\documentstyle[preprint,aps,epsf]{revtex}

\begin{document}
\author{Nara Guisoni\thanks{
Electronic address: nara@fge.if.usp.br} \\
Vera Bohomoletz Henriques\thanks{Eletronic address: vhenriques@if.usp.br} \\ }
\title{Square water as a solvent: Monte Carlo simulations}
\address{Instituto de F\'{\i}sica, Universidade de S\~ao Paulo, \\
C.P. 66318, cep 05315-970, S\~ao Paulo, SP, Brazil}
\maketitle

\begin{abstract}
Square water takes into account the directionality of hydrogen bonds. The model is reviewed and its properties as a solvent for apolar
particles are studied through Monte Carlo simulations. Specific heat
measurements are used to identify phase separation. Data for comparison 
with the lattice gas on the square lattice are presented and the relation to
non-associating solvents is discussed. Data for the frequency of hydrogen
bonds as a function of temperature indicate a slower rate of bond breaking 
for the hydration shell as compared to bulk water particles.
\end{abstract}

\section{introduction}

Apolar molecules are poorly soluble in water \cite{frank evans,pollack}. It has been known for a long time that this is due, in many
cases, not to an enthalpy, but to entropy effects: the dissolution in water
of such molecules results in a decrease both in enthalpy (which would lead
to large solubility) and in entropy, inspite of the disruption of the net of
hydrogen bonds characteristic of water. This effect has been qualitatively
explained quite some time ago \cite{frank evans} in terms of an effect known
as hydrophobic hydration: a tendency of the water molecules to
``strengthen'' their structure around the apolar molecules. Hydrophobic
interaction, leading to a tendency of aggregation of nonpolar solute, would
also result. NMR and other experimental evidence \cite{luck,grigera3}
is in favour of the hydration hypothesis. Molecular dynamics simulations of
model water solutions with atomic detail have probed these ideas, but
evidence is often contradictory \cite{haymet} and still not quite conclusive
on which microscopic properties would be responsible for the lower entropy 
\cite{italianos,madan,urahata,grigera}. The simulation
of molecular models with atomic detail is restricted, by todays' computer
facilities, to short time simulations. Simplified models are therefore an
alternative tool. In a recently proposed very simple lattice model Barkema
and Widom\cite{widom} impose ordering on dissolution and analyse the
effective solute interaction. In this paper we follow the more usual
approach: we adopt a lattice model for the solvent and look for signs of
induced structure around apolar solutes.

The hydrogen bond network\cite{stillinger} is thought to be responsible for
other special properties of pure water such as a maximum in the density
curve as a function of temperature. Beyond condensation the
tetrahedrally coordinated open ice structure is distorted retaining most of
the HB network but accomodating some broken bonds and increasing slightly the
number of nearest neighbours and thus the density \cite{kauzman}. At higher
temperatures, usual expansion takes place, under increasing disrupture of
the net \cite{kauzman}. Consideration of water with atomic detail has led to 
the development of a number of different models \cite{jorgensen,berendsen}, all of which present some sort of discrepancy in relation to experimental
thermodynamic or structural properties of liquid water. On the other hand,
inumerous simplified statistical models \cite{stanley2} have been proposed
in the search for the main features which may explain the behaviour
exhibited by this special liquid. Square water is one of these models, whose
properties were studied by Nadler and Krausche\cite{nadler}. Probably due to
its simple thermodynamic behaviour, in which no phase transition is present,
it is scarcely mentioned in the literature as a model for water \footnote{
Precisely the same model has been used recently to describe ``spin-ice''
behaviour of frustrated ferromagnets\cite{wolf}, but association to Nadler's
square water has not been noticed.}.

Square ice incorporates the hydrogen bond net but does not allow for
distortions, neither for density fluctuations. Thus the model is incapable
of presenting anything like a liquid-ice transition. However, it could
represent the net and its fluctuations under temperature characteristic of
liquid water. Rotations which lead to disrupture of a bond are considered in
an extreme (discrete) form. We have undertaken to study the effect of such
an associated model solvent \cite{egelstaff} on apolar solutes, both in comparison to non
polar solvents, as well as in terms of the possible hydration effect.

The pure and mixture models are defined in the following section. \ Although
presented previouly by Nadler and Krausche square water is explained for
clearness and its connection with ferroeletric vertex models is discussed. \
Simulations and results are presented in Section III and a summary presented
in the final section.

\section{The models and simulations}

\begin{figure}[th]
\epsfysize=5.3 cm \epsfxsize=16.64cm  
\epsffile{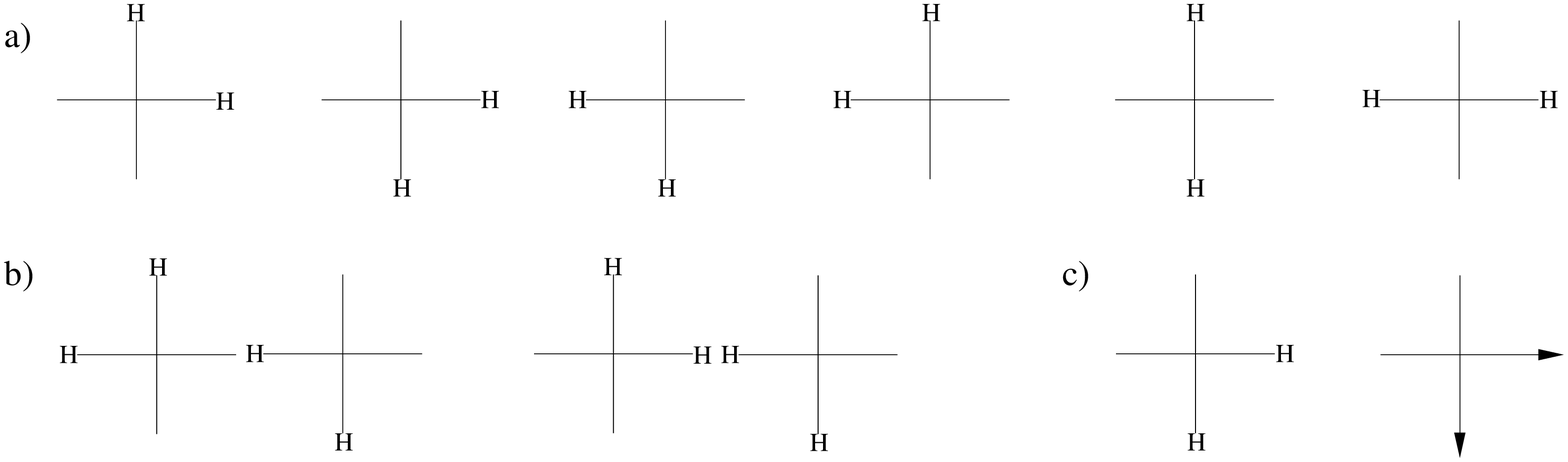}
\vspace*{3cm}
\caption{\footnotesize a) The six states for the water particle. Oxygens are on lattice sites. b) An HB is present if there is one hydrogen atom on the lattice line. The first pair contributes with HB energy $-\epsilon$. Rotation of the left molecule ``breaks'' the bond, as in the second pair. c) Vertex representation of water molecule.
\label{agua}}
\end{figure}

Square water consists of a square lattice whose points are occupied by
oxygens and whose lines \footnote{
In order to maintain a clear distinction between lattice bonds and hydrogen
bonds we have chosen to call the former lattice lines.} are occupied by
hydrogens. There are six different states per water particle, related to the
possible distributions of two molecular hydrogens on four lattice lines
(Fig. \ref{agua}a). The hydrogen bond is present when the hydrogen atom of one
molecule (donor) points to the oxygen of a neighbour one (acceptor), and is
absent when both molecules intend to be donors or acceptors. Typical
situations are illustrated in Fig. \ref{agua}b. In the presence of a hydrogen bond
(HB) an energy $-\epsilon $ is attributed to the lattice line, otherwise the
energy is zero. Because the bond between two neighbour molecules depends on
the relative position of the hydrogen atoms, the model mimics the
directional nature of HBs \cite{stillinger}, despite the fact that it is a
lattice model. \ Hydrogen bonding is the only interaction considered,
since van der Waals interactions which should also be present are an order
of magnitude smaller than HBs \cite{kauzman}.

The model for water is a generalization of square ice \cite{lieb2,nagle} (see Fig \ref{agua}c), through the inclusion of thermal fluctuations which produce
rotations of the molecules. It must not be confused with ice rule
ferroelectric or vertex models \cite{lieb1} for which energy of neighbour
pairs is either zero or infinite (no broken bonds) and different energies
are attributed to the particle states. The present model particles are
allowed six states but the bond graph representation \cite{lieb1} of the
vertices is no longer adequate. As in generalized ferroeletric
models the ice condition is broken, but in a completely different way: each
oxygen atom is always surrounded by two hydrogens near it (representing a
neutral molecule) but it is possible to have two, none or one hydrogen atom
on each lattice line. An HB is present only in the last case. Therefore,
differently from vertex models, the energy depends on the neighbourhood. 
\footnote{
The model could be written in terms of generalized Potts variables with
directional rules for the interactions.}

The hydrophobic properties of a square water solvent are tested on a model
for aqueous solution of apolar molecules. Single-site nonpolar particles are
introduced on the lattice and, for simplicity, are considered inert
(interations with water or among themselves are disregarded). Thus HBs are
allowed to ``break'' due either to the presence of nonpolar particles or, as
in the pure water model, as a result of thermal fluctuation.

The properties of the two models were studied through Monte Carlo
simulations in the canonical ensemble. Two types of movement are needed in
order to go through the phase space of the system. For the pure system, a
new state is obtained using a local movement of the HB network: a water
molecule is randomly chosen and a new state selected. In the mixture system,
with the concentration of nonpolar particles fixed, an additional movement
is introduced which consists of a random distance exchange \cite{shida}
between water and nonpolar particles (a global movement). In the latter
case, the particles are randomly chosen and their spatial positions
exchanged. The state of the water molecule in its new position is chosen
randomly. In both cases the Metropolis probability is used \cite{metropolis}.
Random numbers are generated from ran2 \cite{numerical}.

\section{Results and discussions}

\begin{figure}[th]
\epsfysize=12.77cm \epsfxsize=12.64cm  
\epsffile{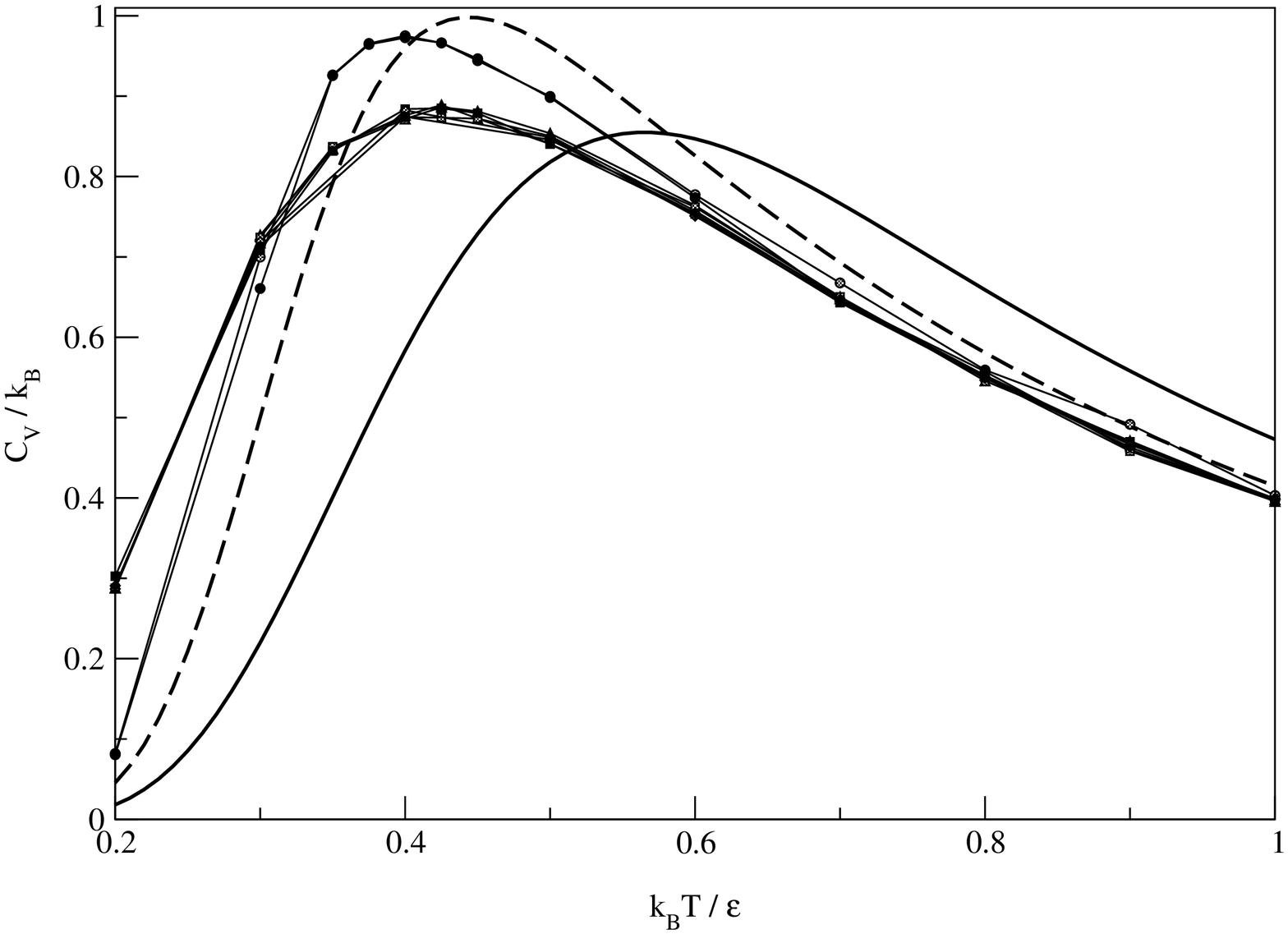}
\vspace*{3cm}
\caption{Specific heat for pure water. Circles, triangles, squares and diamonds correspond to $L=4,10,15,30$ Monte Carlo results, respectively. Grey and black symbols represent ordered and disordered initial conditions. Continuous and dashed lines are exact result for $L=2,3$. Smooth behaviour as a function of system size indicates absence of phase transition.
\label{calor_agua}}
\end{figure}

As shown previously by Nadler and Krausche \cite{nadler} from simulations,
square water does not present a phase transition. In figure Fig. \ref{calor_agua} we
present similar results for several lattice sizes, as well as exact
calculations for small lattices: there are no significant differences in the
specific heats of lattice sizes greater than L=10.\footnote{
This result could be interpreted qualitativaly due to the high
entropy of the ground state (the entropy of the ice model is approximately 1.5$k_{B}$\cite{lieb2}
per particle, and therefore the competition between a state with low energy
and another one with high entropy is smoothed down.)}

\begin{figure}[th]
\epsfysize=12.77cm \epsfxsize=12.64cm  
\epsffile{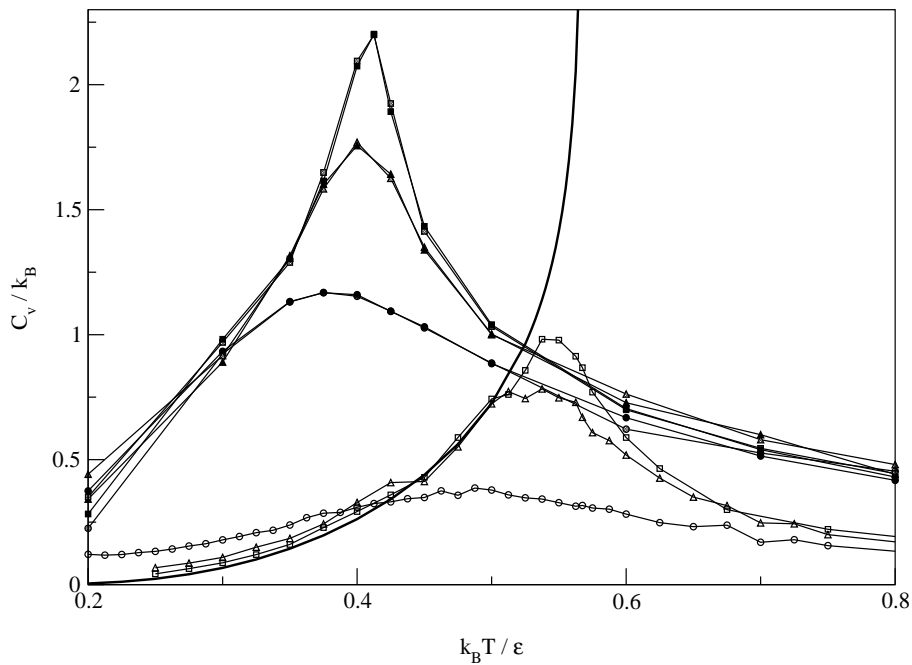}
\vspace*{3cm}
\caption{Specific heat (full symbols) for the solution of apolar particles at $20\%$ concentration. Gray and black symbols represent ordered and disordered intial conditions. Lattice gas specific heat is shown for comparison (empty symbols for MC results and line for exact result). Circles, triangles and squares respectively for $L=10,30,60$. Peaks indicate phase separation at low temperature.
\label{calor_mix}}
\end{figure}

The model solution was studied for concentration 20\%. The system exhibits
phase separation  below a reduced transition temperature $t\equiv
k_{B}T/\epsilon \approx 0.4$ which is indicated by the peak in the specific heat
shown in Fig. \ref{calor_mix}. Lattice gas data at $20\%$ obtained from a random exchange
algorithm \cite{shida2} are shown for comparison, as well as the exact
result \cite{onsager} for the homogeneous phase.

The square water and the \ lattice gas mixtures may be compared respectively
to associated\cite{egelstaff} and van der Waals liquids acting as solvents.
Associated liquids present higher boiling temperature and lower solubility
than corresponding van der Waals liquids. The difference between these two
model solvents is the directionality of the HB interaction. Comparison of
the specific heat peaks show that phase separation occurs for a larger
reduced transition temperature in the case of the lattice gas. In order to
understand this result we must look at the absolute transition temperature.
van der Waals bond energies $\epsilon (vdW)$ are usually an order of
magnitude lower than HB energies $\epsilon (HB)\cite{egelstaff}$. Thus

\begin{equation}
\frac {T_{w}}{T_{vdw}} \approx 0.7 \frac {\epsilon (HB)} {\epsilon (vdW)} \approx 7.
\end{equation}
The absolute transition temperature of the aqueous solution model ($T_{w}$) is\ an
order of magnitude greater than that of the van der Waals solution ($T_{vdW}$),
indicating lower solubility in the first case.  

As to solubility properties of the model, it can be argued that it increases
with temperature, following the tendency of the system to become homogeneous
at higher temperatures\cite{atkins}. It remains to be seen if this
property persists at smaller overall concentrations.

\begin{figure}[th]
\epsfysize=9.77cm \epsfxsize=12.64cm  
\epsffile{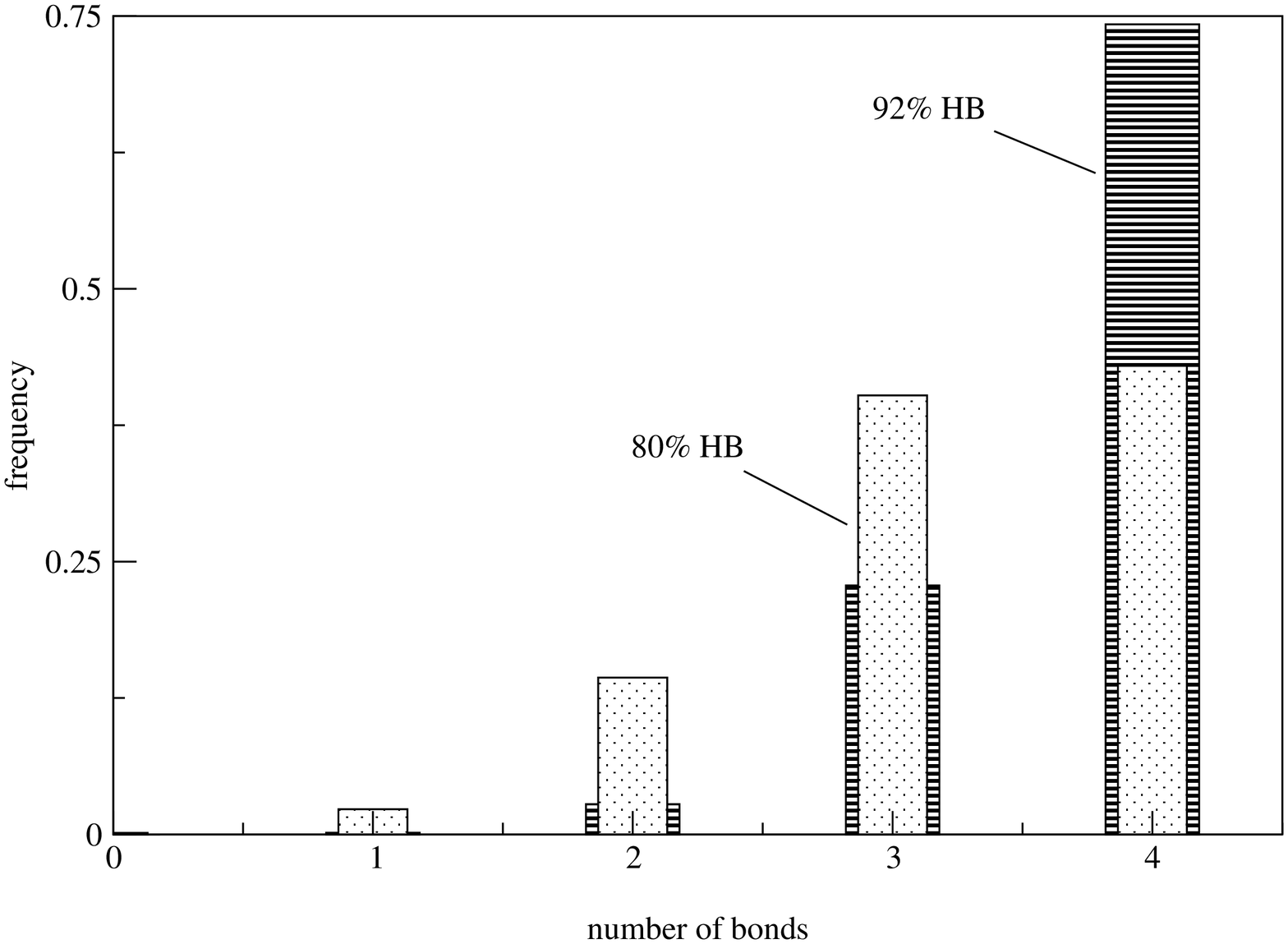}
\vspace*{3cm}
\caption{Pure water: particles classified according to the number of bonds. $80\%$ of bonds are present for reduced temperature $t=0.7$ and $92\%$ of bonds for $t=0.4$. Monte Carlo data for $L=60$.
\label{tipos}}
\end{figure}

The study of water structure was performed in terms of frequency of bonds.
At ambient temperatures the liquid water net presents around 88\% of maximum
number of bonds \cite{jorgensen}. In Fig. \ref{tipos} we present a histogram of
particles classified according to the number of bonds (there are between 0
and 4 HBs per particle) for pure square water typical of this region (80 and
92\% of HB, the latter corresponding to a temperature slightly below
transition). 

\begin{figure}[th]
\epsfysize=12.77cm \epsfxsize=12.64cm  
\epsffile{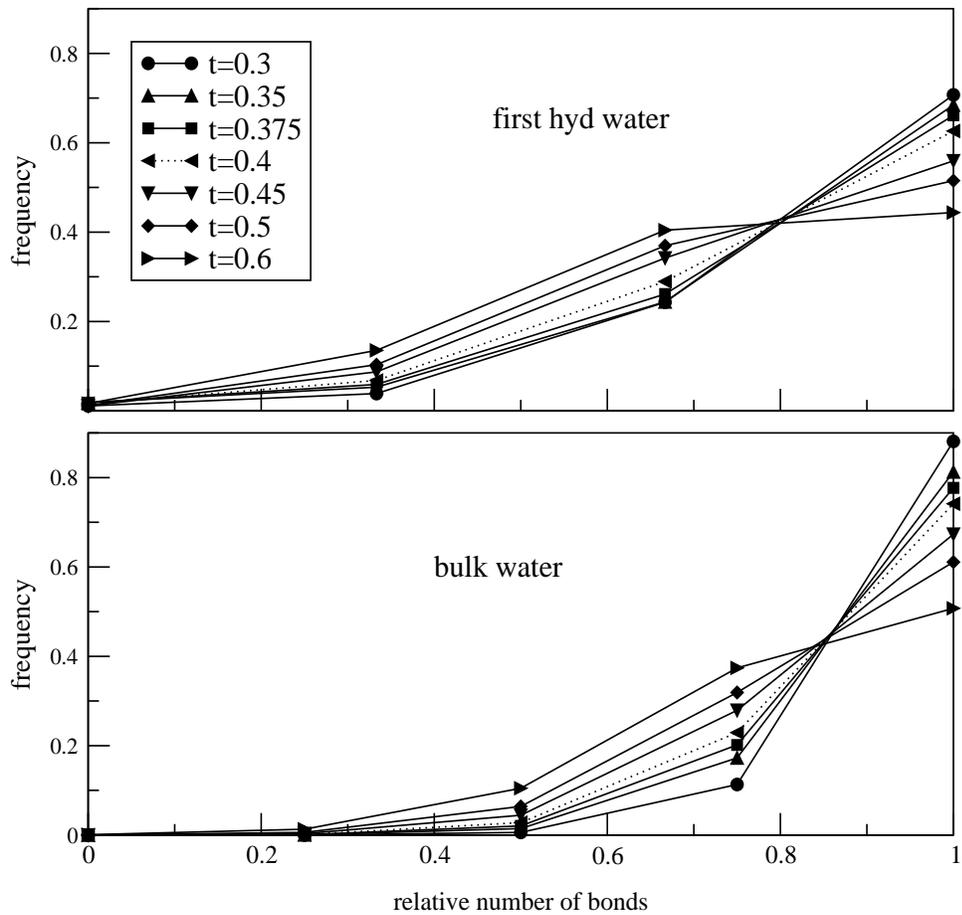}
\vspace*{3cm}
\caption{First hydration shell and bulk water particles classified according to the number of bonds. Dashed line is phase separation temperature. Note slower rate of variation with temperature for hydration shell particles both above and bellow transition temperature. Monte Carlo data for $L=60$.
\label{tipos_mix}}
\end{figure}

For the mixture we have classified separately first hydration shell (first neighbours of apolar particles) and bulk
water particles. The water molecules around the nonpolar particles make less
bonds and have a smaller frequency for maximum bond number. This result is
in agreement with that obtained from simulations of atomic detail solvent 
\cite{urahata}. On the other hand, frequency of maximum bond number in the first hydration shell shows a smaller dependence on temperature as
compared to bulk water, as can be seen in Fig. \ref{tipos_mix}. The hydration shell shows
slower rate of decay of bonds with temperature above and specially below the phase separation
transition. Thus the ratio of frequencies of maximum bond number for
shell vs bulk water increases with temperature. Similar results have been
seen in simulations of atomic detailed water\cite{urahata,grigera},
albeit in contradiction to the simplified off-lattice MB model\cite{dill}.

\section{ Conclusions}

We have compared directional bonded (square water) and simple-bonded (lattice
gas) solvents for apolar solutes in relation to phase-separation in terms of their specific
heats. For 20\% concentration square water presents lower reduced
transition temperature but the associated HB net which it is intended to
described presents a higher transition absolute temperature. At the
concentration studied solubility increases with temperature contrary to what
one would expect for apolar solutes in water. Simulations at different
concentrations must be undertaken in order to check whether this is an
overall property of the model.

Hydrophobic properties of square water were also studied in terms of
frequency of bonds of bulk and hydration shell water. Hydration shell
water ``looses'' structure with temperature at a slower rate than bulk
water. This might contain some indication of the presence of
``hydrophobic'' hydration \cite{frank evans}.

The study of the effect of such properties on aggregation or structure of
amphiphilic molecules may be of interest.

\section{Acknowledgments}
We thank P. Wolf for indicating  relation to the ``spin ice'' problem and
C.E.I. Carneiro for discussions of simulations. One of us (NG) aknowleges
financial support from Fapesp.

\newpage

\newpage

\newpage

\newpage

\vspace*{3cm}

\newpage

\end{document}